\documentclass[preprint]{revtex4}
\usepackage{graphicx}
\usepackage{dcolumn}
\usepackage{bm}
\usepackage{amsmath}
\usepackage{verbatim}

\begin{document}

\title{ A Theory of Loop Formation and Elimination By Spike Timing-Dependent Plasticity  }

\author{James Kozloski}
\author{Guillermo A. Cecchi}
\affiliation{Computational Biology Center, IBM Research Division,
IBM T.J. Watson Research Center, Yorktown Heights, NY 10598}

\date{\today}

\begin{abstract} (FULL PAPER IN FRONTIERS IN NEURAL CIRCUITS, 2010) We show that the local Spike Timing-Dependent
Plasticity (STDP) rule has the effect of regulating the
trans-synaptic weights of loops of any length within a simulated
network of neurons. We show that depending on STDP's polarity,
functional loops are formed or eliminated in networks driven to
normal spiking conditions by random, partially correlated inputs,
where functional loops comprise weights that exceed a non-zero
threshold. We further prove that STDP is a form of loop-regulating
plasticity for the case of a linear network comprising random
weights drawn from certain distributions. Thus a notable local
synaptic learning rule makes a specific prediction about synapses in
the brain in which standard STDP is present: that under normal
spiking conditions, they should participate in predominantly
feed-forward connections at all scales. Our model implies that any
deviations from this prediction would require a substantial
modification to the hypothesized role for standard STDP. Given its
widespread occurrence in the brain, we predict that STDP could also
regulate long range synaptic loops among individual neurons across
all brain scales, up to, and including, the scale of global brain
network topology.
\end{abstract}

\maketitle

\section{Introduction}

Connections between individual neurons in the brain are constrained
first by the spatial distribution of axons and dendrites within the
neuropil \cite{BRAITENBERG}\cite{CHKLOVSKII}. Global brain networks
comprise  dense connections within tissues, the gross structures in
which these tissues are embedded, and the bidirectional long-range
projections joining these structures. The topology of these networks
is not yet fully specified at the level of microcircuitry, however
\footnote{We use microcircuitry to refer to neural circuitry
observed at the level of neurons and synapses and not necessarily in
reference to a restricted spatial extent of these neurons (e.g., a
``column''). For this reason, we view every brain connection as part
of some microcircuit topology.}. One theoretical constraint on this
level of organization, the ``no strong loops hypothesis,''
considered only developmentally determined area to area connectivity
patterns to implement its specific neuron to neuron network
topological constraint \cite{CRICK}. While local synaptic
modifications are known to directly shape the pattern of
connectivity in local neural tissue and thus local microcircuit
topology \cite{MARKRAM06}, our understanding of global brain network
topology still derives largely from this developmentally patterned,
area to area connectivity. Furthermore, measuring simultaneously the
relative strengths of specific microcircuit connections remains
technically challenging, and virtually impossible for even medium
sized ($100-200$ neurons, $0.05-0.1$ mm) microcircuits. For these
reasons, it is not yet known how large scale, long range
microcircuit topology and the computation it supports emerges
through synaptic modifications in the brain.

We wondered whether a synaptic modification commonly observed in
local circuit preparations and widely hypothesized to shape local
dynamics in brain structures, STDP~\cite{MARKRAM97}, could be
analyzed to yield an understanding of what topology it predicts for
microcircuits of any scale. The STDP model is a departure from
traditional Hebbian models of learning, which state that neurons
that fire action potentials together will have their
interconnections strengthened. Instead, STDP takes into account the
particular temporal order of pre- and post-synaptic neuronal firing
\cite{MORRISON08}, such that the rule modifies synapses
anti-symmetrically, depending on whether the pre- or post-synaptic
neuron fires first (Fig. 1). The basic question we then aimed to
answer is: what is the influence of this anti-symmetry on brain
microcircuit topology?

Consider first if a pre-synaptic ``trigger'' neuron causes a
post-synaptic, first-order ``follower'' neuron to fire. If this
follower makes a direct feedback connection onto the trigger, the
feedback connection will be weakened, since the spike generated by
the follower will arrive at the follower-trigger synapse immediately
after the trigger neuron's backward propagating action potential
(Fig. 1). The principle that STDP is suitable for eliminating strong
recurrent connections between two neurons was originally proposed by
Abbott and Nelson \cite{ABBOTT_NELSON}. Here we expand on the
principle with the observation that it holds for \emph{all}
polysynaptic loops connecting triggers and followers: if some
$n^{th}$-order follower's action potential produces in the original
trigger a subthreshold potential after the trigger has fired, the
functional loop will be broken by spike-timing dependent synaptic
weakening of the feedback connection. With this intuition, we set
out to prove analytically and by means of numerical simulation that
network topology, and specifically the occurrence of functional
loops in highly connected networks, is directly and necessarily
regulated by STDP.

This theory paper provides clear predictions about STDP's effect on
neural circuit topology. The proof and simulations dictate strong
constraints on local and long range microcircuit connectivity. We
propose that if these constraints are not obeyed by real neural
circuits, the hypothesis that standard STDP shapes the structure and
function of real nervous systems must be revised. Our approach
suggests that similar analyses of other learning rules may impose
similar constraints on neural circuit topology and that the
hypothetical significance of these rules may similarly be tested.

\section{Results}
\subsection{A proof of STDP as a form of loop-regulating plasticity}

First, we represent STDP acting on a weight $w$ associated with the
connection between two neurons and their output variables $x(t)$ to
$y(t)$, in the adiabatic approximation (i.e. small learning rate),
as:
\begin{equation}\label{eqn:1}
\Delta w_{xy} \propto \int_{-\infty}^{\infty} C_{xy}(t) S(t) dt
\end{equation}
where $C_{xy}(t) = \int x(t^{\prime}- t) y(t^{\prime})$ is the
correlator, and $S(t)$ is the anti-symmetric STDP update function,
$S(t<0)=\exp(\lambda t)$, $S(t>0 )=-\exp(-\lambda t)$. Consider this
function operating over connections within a linear network driven
by uncorrelated Gaussian inputs, $\xi$, such that $\dot{x}(t) = W
x(t)+\xi(t)$, where $x$ is a vector of activities with components
denoted by $x_i$, the weight connection matrix has components
$W_{ij}$, and the input satisfies $\langle \xi_i(t)
\xi_j^{T}(t+\tau)\rangle_t= \sigma^2 \delta(\tau) \delta_{ij}$. We
show (see Appendix \ref{SuppInf:Update}) that the learning rule
defined in Eq.~\ref{eqn:1} results in an update for the network
weight matrix of the form $\Delta W=\Delta W(W,\tau,C_0)$ where
$\tau$ is the time constant of the STDP's exponential, and $C_0$ is
the instantaneous correlator $C(0)$. This update rule influences
global network topology in a very specific way.

To formalize our original intuition analytically, consider a linear
network with only excitatory connections, such that the dynamics may
be expressed as ${\dot x}= W x=(-I+A)x$, where $A_{ij} \geq 0$ is
the network connectivity matrix (comprising the off-diagonal
elements of the weight matrix and zeros on the diagonal), and $-I$
represents a self-decay term. Next, we introduce a ``loopiness''
measure that estimates the strength of all loops of all sizes that
occur in the network, $\mathcal{E}_l=\sum_{k=1}^{\infty} \frac{1}
{k} \text{tr}~ \left[A^{k}\right]$. The function
$\text{tr}~\left[\cdot\right]$ stands for the trace operation; this
operation, when acting on the $k$-exponentiation of the adjacency
matrix (comprising ones and zeros, where the nonzero entry $a_{ij}$
represents a connection from network node $i$ to network node $j$),
counts the total number of closed paths of length exactly equal to
$k$, i.e. $k$-loops \footnote{Note that paths that traverse the same
network node more than once are also counted.}. When applied to the
network connectivity matrix $A$, the operation counts loops weighted
by the product of the synaptic strengths of the looping connections,
resulting in a slightly different, but still useful, measure of
loopiness.

It is possible, however, to reduce this measure without actually
regulating topology by simply reducing the weights of all
connections. A \emph{topological} loopiness measure should therefore
include a penalty to the weights' vanishing; we choose
$-\frac{1}{2}\text{tr}~\left[A A^{T}\right]$, ($^{T}$ stands for the
transpose operation) which, for weighted graphs, measures the sum of
the squares of all network weights (and for a binary graph, counts
the number of links). We then define the total topological loopiness
as:
\begin{equation}\label{eqn:3}
\mathcal{E} = \sum_{k=1}^{\infty} \frac{1} {k} \text{tr}~
\left[(A)^{k}\right] - \frac{1}{2}\text{tr}~ \left[A A^{T}\right]
\end{equation}
We showed computationally that for every weight matrix examined,
each drawn from certain random distributions (see Appendix
\ref{SuppInf:Update}), the change in this energy as a function of
the evolution of the network under STDP, $\Delta \mathcal{E} \sim
\text{tr}~\left[\partial_A \mathcal{E} \Delta A^{T}\right]$, is
strictly semi-negative, and therefore STDP necessarily regulates
this measure, resulting in a decrease in topological loopiness. We
therefore use the term STDP and ``loop-regulating plasticity''
interchangeably throughout.

\subsection{Loop-regulating plasticity in a network of simulated neurons}
What are the effects of this form of plasticity on network topology
(and specifically on the number of functional loops) in nonlinear
networks, such as those found in neural microcircuits? Because our
proof of STDP as a form of loop-regulating plasticity applies only
to linear networks or nonlinear networks that may be linearized, we
aimed to show, using simulation, that the same principle extends to
a biologically relevant, nonlinear regime. We replicated the
simulation of Song and Abbott \cite{SONG}, extending it in three
ways (see \ref{Simulation Methods}). First, we created a network of
100 neurons, each receiving excitatory synapses from all other 99
``intra-network'' input sources and from 401 randomly spiking
``extra-network'' input sources selected at random from 2,500
homogeneous Poisson processes. All excitatory synapses underwent
STDP. Second, we provided 250 inhibitory synapses to each neuron,
sampled from 1,250 spiking sources; the inhibitory inputs modeled
fast local inhibition to the network using inhomogeneous Poisson
processes with rates modulated by the instantaneous aggregate firing
rate of the network. Third, we explored four different forms of STDP
update \cite{BURKITT04} and observed robust loop-regulating
plasticity for each; the results presented here used the STDP update
rule of G\"{u}tig et al. \cite{GUTIG}.

We initialized our network with maximum extra-network weights, and
intra-network weights at half maximum. This caused the network to
spike vigorously when extra-network inputs became active, but spike
rates were limited by the fast local-inhibition. After 20 seconds of
simulated network activity, STDP had a profound effect on
topological loopiness as defined in Eq.~\ref{eqn:3}, measured over
loops of length $2 \leq k \leq 100$ for convenience (Fig. 2A). We
counted the number of closed, functional loops of varying length
using $\text{tr}~\left[\lceil A\rceil^{k}\right]$, where $\lceil A
\rceil$ was constructed by applying a sliding threshold to the
network connectivity matrix (Fig. 2B). We compared this quantity to
the same, measured for a randomized network, constructed by randomly
reassigning weights from the learned weight distribution to synapses
in the network (see \ref{Analysis Methods}). These results are
representative of all loop lengths measured ($2\leq n\leq 100$) and
show that as the weight threshold grows, the number of closed,
functional loops in the STDP-learned network decreases more than in
the randomized network. This form of loop-regulating plasticity can
therefore be described as loop-eliminating.

\subsection{The effect of synaptic delays on loop-regulating plasticity}

We wondered what effect synaptic delays would have on this result,
since we expected follower feedback spikes to cause less anti-loop
learning as they fell further from the zero time difference maxima
in the STDP update function. We also wondered if the decrease in the
number of functional loops compared to a randomized network also
applied to unique loops, in which no neuron is traversed more than
once. We therefore sampled the number of unique, functional loops
through networks simulated with synaptic delays from $0.1$ to $4.0$
milliseconds. We constructed one million random paths of length
$k-1$ for each loop length $2 \leq k \leq 25$, and for the learned
and randomized networks (see \ref{Analysis Methods}). We searched
for each path across all networks studied, and if the path and the
$k^{th}$ link completing the functional loop existed in the network,
we counted it for that network (see \ref{Analysis Methods}). The
result is similar to that for closed loops, and, as expected, longer
synaptic delays resulted in an exponential decrease in the number of
loops as a function of loop length that deviated less from the same
function for randomized networks, indicating weaker loop-regulating
plasticity (Fig. 2C).

\subsection{Network in-hubs, out-hubs, and loop-regulating plasticity}

Next, we asked if other topological measures of the STDP-learned
networks may be correlated with our observation of STDP's effect on
loopiness, since many different topological properties might
coincide with or support this effect. For example, one means to
create networks poor in loops is to ensure that nodes in the network
are either ``out-hubs'' or ``in-hubs,'' but not both \cite{CYCLES}.
An out-hub in a network of neurons has many strong postsynaptic
connections but few strong presynaptic connections, and an in-hub
has many strong presynaptic connections but few strong postsynaptic
connections. We applied a sliding threshold to the network
connectivity matrix learned by STDP, and examined the manifold,
colored according to each applied threshold, which correlated
in-degree versus out-degree for each neuron in our network. This
showed a clear inverse relationship between in- and out-degrees that
varied in form with weight threshold (Fig. 3A). In contrast, by
examining the in-degree from extra-network inputs, we found a
positive correlation (Fig. 3B), indicating that out-hubs were more
likely to be in-hubs within the larger extra-network topology, and
that in-hubs in our network were more likely to receive only the
weakest extra-network inputs.

\subsection{Reverse STDP restored loops after loop-eliminating plasticity}

Beyond these standard topological analyses, we also examined
biological properties of the network. We measured total synaptic
input as a function of total synaptic output for all neurons in the
STDP-learned network. In the same experiment, we asked if reversing
the polarity of the standard STDP function might undo the effects of
loop-regulating plasticity that results from standard STDP, since
under this ``reverse'' condition, follower spikes would cause
strengthening of closed-loop feedback connections. This reversal of
polarity is biologically relevant, since it occurs at the synaptic
interface between major brain structures such as neocortex and
striatum \cite{STDP-reverse2}, arises specifically at synapses
between certain cell types, and is controlled by cholinergic and
adrenergic neuromodulation, for example in the neocortical
microcircuit \cite{STDP-reverse1}. We found the same inverse
relationship between in-degree versus out-degree for each neuron in
our network (Fig. 4A, green markers), as well as an inverse
relationship between total synaptic input and output following $1.5$
seconds of standard STDP (Fig. 4B, green markers). These effects
contributed to a reduction in the number of closed loops (Fig. 4C,
depicted as in Fig. 2B), and each of these relationships could be
largely abolished by $3$ to $5$ additional seconds of reverse STDP
(Fig. 4, red markers), in contrast to $3$ to $5$ additional seconds
of standard STDP (Fig. 4, blue markers), which strengthened them. We
also found the same positive correlation between in-degree from
extra-network inputs and out-degree within the network (Fig. 4D) and
between total extra-network synaptic input and total intra-network
synaptic output (Fig. 4E). This effect was also largely abolished by
$3$ to $5$ seconds of reverse loop-regulating plasticity, but
reinforced by $3$ to $5$ seconds of standard loop-regulating
plasticity.

\subsection{Dynamical effects of loop-regulating plasticity}

What are the consequences of this form of network plasticity beyond
topology? In the case of a linear network, reducing the number of
loops implies more stable dynamics. Consider the stability of the
unforced system $\dot{\vec{x}}(t) -  W \vec{x}(t)=0$; the
eigenvalues $\lambda$ of $ W =- I + A $ can be expressed as:
\begin{equation}\label{EigLinear}
- \sum_{i=1}^N \log|\lambda_i| = \sum_{k=1}^{\infty} \frac{1}{k}
\text{tr}~\left[ A^k\right]
\end{equation}
\cite{PRASOLOV}, which emphasizes the contribution of loops to
system instability. Such a simple observation, however, does not
make clear predictions about the effects of loop-regulating
plasticity on nonlinear neural circuit function. We were surprised
to find that raster plots of network spiking activity, when sorted
according to certain topological metrics (e.g., the sum of
extra-network input weights, the sum of intra-network output
weights, in-degrees, out-degree) consistently revealed network
events that originate with weak synchronization among out-hubs,
followed by strong synchronization among in-hubs (Fig. 5A, top),
across $8$ independent simulations of the phenomenon. This effect
was altered by randomizing the intra-network weights, such that
synchronization events became stronger, more frequently global, and
more frequent among out-hubs alone (Fig. 5A, bottom). Peri-event
time histograms constructed across $8$ independent simulations
reveal this same effect (Fig. 5B, left panels), with synchronization
arising strongly among in-hubs after weak out-hub activation in the
STDP-learned network, and globally in the randomized network (see
\ref{Analysis Methods} for a description of how network events were
detected). In the STDP-learned network, both in-hubs and out-hubs
sustain spike rates ranging from $4-9$ Hz that are not correlated
with in-degree, whereas in the randomized network, spike rates range
more broadly ($3-16$ Hz) and are highly correlated with in-degree
(Fig. 5B, right panels). We examined the summed network peri-event
time histograms for the STDP-learned network and for networks that
underwent randomization of their intra-network weights, their
extra-network weights, or both (Fig. 5C, top), across $8$
simulations for each condition. The resulting distributions, as well
as a pooled distribution of times from all randomized networks each
differed from each other based on paired Kolmogorov-Smirnov tests
($P\approx 0$). To examine what properties of these distributions
distinguished them, we measured kurtosis and skew for each
distribution from each simulation, and compared the distributions of
kurtosis and skew measures between each group. Kurtosis differed
significantly between the STDP-learned network topology and the
randomized topologies (intra-network randomized, extra-network
randomized, both randomized, Fig. 5C, bottom). Results from unpaired
t-tests of the distribution of skew and kurtosis measurements
between $4$ simulation conditions (calculated separately for each of
the $8$ simulations) are shown in table inset (Fig. 5C, bottom).
These effects indicate that the effect of standard loop-regulating
plasticity is to generate network topologies that support network
events with greater spread and sharper peaks in time.

\section{Discussion}
Based on our simulations and analytical results, we propose that
standard STDP must produce a network topology in real neural tissues
that is conspicuously poor in both closed and unique loops, and that
it will segregate neurons into out- and in-hubs to achieve this.
Such a prediction can easily be tested by analyzing correlations
between the number of functional input connections and the number of
functional output connections made by neurons recorded during
multi-patch clamp experiments in a structure in which STDP has been
observed (e.g., \cite{MARKRAM06}, \cite{SONG_CHKLOVSKII}). Our
theory predicts this correlation should be negative.

The network that emerges in such tissues will organize its
relationship to inputs from other structures in an orderly fashion,
making local out-hubs the primary target for long range inputs, and
thus establishing a feed forward relationship between the network
and its pool of inputs. In a larger system, we anticipate that local
in-hubs would become long range outputs. This prediction may also be
tested by correlating the local topological relationships of a
neuron with its identified role as either an input, output, or
interneuron within that structure.

We also make a clear prediction for the effect of synaptic delays on
modulating the topological effects of STDP. Correlations between
these delays and functional connectivity data from multi-patch clamp
recordings from connected neurons undergoing STDP can also be
measured to determine if synaptic delays predict the strength of
reciprocal connections. Furthermore, we observe that, at interfaces
between brain structures where STDP is reversed by neuromodulation,
circuit dynamics can be predicted from the expected change in
network topology. For example, changes in STDP at the
cortico-striatal synapse \cite{STDP-reverse3} resulting in reverse
STDP would favor the emergence of strong
cortico-striatal-thalamocortical loops resulting in oscillations in
this circuit.

Interestingly, the depletion of loops and the separation of nodes
into out-hubs and in-hubs has been recently reported in a variety of
complex biological systems, including functional networks at the
level of spatio-temporal resolution of fMRI, and the neural network
of C. Elegans \cite{CYCLES}, suggesting a general principle of
organization and dynamical stability for entire classes of
functional networks. These observations suggest quantitative
measurements of topology in vertebrate microcircuits could produce
similarly interesting results.

It has been observed in local circuit preparations that a bias
exists among layer 5 pyramidal neurons of rat neocortex towards
strong reciprocal connectivity \cite{MARKRAM06}
\cite{SONG_CHKLOVSKII}, and towards looping motifs among triplets of
this neuronal class \cite{SONG_CHKLOVSKII}. Furthermore, cyclic
connections are strongest among those neurons connected by the
strongest synaptic weights. These same neurons also exhibits STDP at
the excitatory synapses that join them \cite{MARKRAM97}. Given our
analysis, it is now clear that these observations contradict each
other; specifically, we have shown that standard STDP under normal
spiking conditions with random uncorrelated inputs is
loop-eliminating. Therefore, other mechanisms and constraints than
those we have analyzed must be at play.

Consider the case of networks that have recently spiked at abnormal
rates, either due to increased excitability within the network
(e.g., due to injury, epilepsy, etc.) or due to otherwise elevated
extra-network inputs. If the majority of post-synaptic potentials
immediately cause action potentials, standard STDP may have the
effect of strengthening loops. Also, a network driven by highly and
specifically correlated inputs may spike in temporal patterns
conducive to loop-strengthening by STDP (a hypothesis we are
currently studying). Finally, as we have shown, a spiking network
that has recently experienced a reversal of the polarity of STDP
will also show an increase in the number of loops observed. Clearly
more experiments and observation would be required in order to
confirm or rule out each of these mechanisms.

We observe that network activity propagates smoothly through the
feed-forward topology generated by STDP (Fig. 5A, top panel) without
segregating neurons by average spike-rates (Fig. 5B, upper right
panel). The effect on global brain function of such properties would
include stable average firing rates shared among all neurons,
regardless of their topological position, and robust signal
propagation, similar to ``synfire chains'' \cite{ABELES, HOSAKA}.
Finally, our theory holds that the reversal of STDP's polarity
represents a local switch for the modification of both global brain
network topology and global brain dynamics. Thus sources of
modulation~\cite{Dopamine} that accomplish this reversal locally are
in fact regulating global brain function by means of this switch.

\section{Methods}\label{Methods}
\subsection{Simulation}\label{Simulation Methods}
The simulation methods of Song and Abbot \cite{SONG} were used to
simulate each neuron in our $100$ neuron network. We observed the
reported topological results in each simulation after $10$ seconds
of network activity. In some cases, we performed additional, longer
simulations of network activity to explore the convergence and
stability of these topological measures under various conditions.
Briefly, each neuron model was integrate-and-fire, with membrane
potential determined as in \cite{SONG}, by $\tau_m
\frac{dV}{dt}=V_{rest}-V+g_{exc}(t)(E_{exc}-V)+g_{inh}(t)(E_{inh}-V)$,
with $\tau_m=20\text{ms}$, $V_{rest}=-60\text{mV}$,
$E_{exc}=0\text{mV}$, $E_{inh}=-70\text{mV}$,
$V_{thresh}=-54\text{mV}$, and $V_{reset}=-60\text{mV}$.

The synaptic conductances $g_{exc}$ and $g_{inh}$ were modified by
the arrival of a presynaptic spike, as in \cite{SONG}, such that
$g_{exc}(t)\rightarrow g_{exc}(t)+\bar{g}_a$, and
$g_{inh}(t)\rightarrow g_{exc}(t)+\bar{g}_{inh}$. In the absence of
a spike, these quantities decay by $\tau_{exc}
\frac{dg_{exc}}{dt}=-g_{exc}$, and $\tau_{inh}
\frac{dg_{inh}}{dt}=-g_{inh}$, with $\tau_{exc}=\tau_{in}=5.0$,
$\bar{g}_{in}=0.015$, $0\leq\bar{g}_a\leq\bar{g}_{max}$, and
$\bar{g}_{max}=0.01$. We initialized all elements $\bar{g}_a$ to
different values for intra-network ($\bar{g}_a=0.005$) and
extra-network ($\bar{g}_a=0.01$) inputs.

For extra-network inputs, excitatory homogeneous Poisson spike
trains were generated at a rate of 20 Hz. Inhibitory, inhomogeneous
Poisson spike trains that model fast local inhibition were generated
at a rate $r_{min}\leq r_{inh}\leq r_{max}$, where $r_{min}=5$ Hz,
and $r_{max}=1000$ Hz. On each time step, $dt=0.1$ ms, $r_{inh}$ is
incremented by an amount proportional to the fraction, $\gamma$, of
network neurons that spiked during that timestep, and decays with a
time constant, $\tau_r$, such that
$\tau_r\frac{dr_{inh}}{dt}=-\left[r_{inh}+(r_{max}-r_{min})\gamma\right]$.
After this update, if $r_{inh}$ exceeds $r_{max}$, $r_{inh}
\rightarrow r_{max}$.

For all simulations we report here, the STDP update rule for a
synapse from neuron $j$ to neuron $i$ was $\bar{g}_a(i,j) =
\bar{g}_a(i,j) + \bar{g}_a(i,j)^\mu M(i)$ for synaptic potentiation,
and $\bar{g}_a(i,j) = \bar{g}_a(i,j) +
(\bar{g}_{max}-\bar{g}_a(i,j))^\mu Pa(i,j)$, for synaptic
depression, $\mu=0.1$ ($\bar{g}_a$ is maintained in the interval
$\left[\bar{g}_{min}, \bar{g}_{max}\right]$). As in \cite{SONG},
$M(i)$ and $P(i,j)$ decay exponentially, such that
$\tau_{-}\frac{dM}{dt}=-M(i)$ and $\tau_{+}\frac{dP_a}{dt}=-P_a$,
$\tau_{+}=\tau_{-}=20$. Also as in \cite {SONG}, $M(i)$ is
decremented by $A_{-}$ every time a neuron $i$ generates an action
potential, $A_{-}=0.00035$, and $P_a(i,j)$ is incremented by $A_{+}$
every time a synapse onto neuron $i$ from neuron $j$ receives an
action potential, $A_{+}=0.00035$. This update rule effectively
implements the asymmetric function of STDP (see Fig. 1).

\subsection{Analysis}\label{Analysis Methods}
To randomize our networks for analysis (Figs. 2B, 2C, 4C, 5), we
created a random sequence of indices ranging uniquely from $1$ to
$n$, where $n$ was the number of off diagonal elements in our
network's weight matrix. We used these indices to shuffle uniquely
the positions of all off diagonal elements in the matrix, thus
preserving the network's learned weight matrix, while destroying its
learned topology.

We chose to sample unique loops (Fig. 2C) in the networks rather
than enumerating them, since for long loops ($k>20$) the number of
possible paths to search would exceed $10^{20}$. We therefore
constructed one million random paths of length $k-1$ for each loop
length. We term these paths ``unique'' because we sampled units
within each without substitution (i.e., no sub-loops were sampled in
which a single node is traversed more than once). However, we
allowed that each path could be represented more than once in the
one million constructed paths. In fact, for the shortest paths
constructed ($k\leq 3$) this was necessarily the case since the
total number of possible unique paths is less than one million.

We sampled unique loops from adjacency matrices constructed such
that the weight threshold produced a matrix that was precisely
half-full (for $100$ neurons, a matrix with $5,000$ ones and $5,000$
zeros). For $4.0$, $2.0$, $1.0$, $0.5$, and $0.1$ millisecond
delays, these thresholds were $0.0032$, $0.0030$, $0.0033$,
$0.0037$, and $0.0046$. In this way, we controlled across
experiments for varying weight distributions, and sampled the same
number of links for loops across all experiments. Counts were
compared against the randomized network constructed from $5,000$
links.

To detect network events (Fig. 5B, 5C), we employed the method
described by Thivierge and Cisek \cite{THIVIERGE}. Briefly, for each
simulation in which network events were detected, we generated spike
trains at $1$ msec resolution for each neuron, equal in duration to
the time series analyzed and comprising the same number of spikes as
observed for that neuron. We constructed network spike time
histograms of these spike trains across all neurons using a bin
width of $10$ msec. We then determined a threshold for the network
equal to a count which $5$\% of these bins exceeded. Thresholds were
determined $1000$ times for each network and each simulation, and
the mean of these $1000$ values used as the threshold above which
network events were detected in network spike time histograms for
each simulation.

\newpage
\appendix

\section{Supplementary Information}

\subsection{Update rule for the weight matrix}\label{SuppInf:Update}

The classical definition of Hebbian learning for the weight $w$
connecting two dynamical variables $x(t)$ and $y(t)$ can be written,
in its simplest form, as:
\begin{equation}\label{eqn:HebbNoTime}
\Delta W = \eta C_{xy}
\end{equation}
\begin{equation}\label{eqn:CorrZeroLag}
C_{xy} = \int_{-\infty}^{\infty} x(t)y(t)dt
\end{equation}
Where $\eta$ is the learning constant, which for exposition's sake
will be set to 1. It is important, however, to keep in mind that in
order to write Eqs.~\ref{eqn:HebbNoTime}-\ref{eqn:CorrZeroLag} we
are assuming an adiabatic approximation, i.e. the learning is small
enough that the system can be considered to be in steady-state for
the purpose of computing the correlation.

A natural extension of
Eqs.~\ref{eqn:HebbNoTime}-\ref{eqn:CorrZeroLag} is to introduce {\sl
time}, i.e. to consider delayed as well as instantaneous
correlations:
\begin{equation}\label{eqn:HebbTime}
\Delta W \sim \int_{-\infty}^{\infty} C_{xy}(t) S(t) dt
\end{equation}
\begin{equation}\label{eqn:CorrLag}
C_{xy}(t) = \int_{-\infty}^{\infty} x(t^{\prime}- t) y(t^{\prime})
dt^{\prime}
\end{equation}
It is assumed that the time-dependent weight function vanishes for
long delays, $\lim_{t\rightarrow \pm \infty} S(t) = 0$; the
classical learning rule is recovered when $S(t)=\delta(t)$. If, as
experimental results describing STDP strongly suggest
\cite{MARKRAM97}, the weight function displays strict temporal
anti-symmetry, i.e. $S(t)=-S(-t)$, then
\begin{equation}\label{eqn:HebbSplit1}
\Delta W \sim \int_{-\infty}^{0} C_{xy}(t)S(t)dt+\int_{0}^{\infty}
C_{xy}(t)S(t)dt
\end{equation}
\begin{equation}\label{eqn:HebbSplit2}
\Delta W \sim
\int_{0}^{\infty}\left[C_{xy}(t)-C_{xy}(-t)\right]S(t)dt
\end{equation}

A multi-dimensional linear system driven by uncorrelated input can
be described as:
\begin{equation}\label{eqn:stoch_dynamics}
\dot{x}(t) = Wx(t) + \xi(t)
\end{equation}
where each unit is independently subject to Gaussian white noise
$\xi(t)$, a vector whose components satisfy $\langle \xi_i(t)
\xi_j(s) \rangle = \sigma^2 \delta_{ij} \delta(t-s)$. The lagged
correlator is related to the zero-lagged correlator by
\cite{RISKEN}:

\begin{eqnarray}\label{eqn:CorrNetw2}
C(t) = \left\{ \begin{array}{l l}
e^{W |t|}C_0 & \quad \mbox{$t<0$}\\
C_0 e^{W^{T} t} & \quad \mbox{$t>0$}\\
\end{array} \right.
\end{eqnarray}
where for notational convenience we name $C_0 = \int x^T(t) x(t)
dt$, i.e. the correlator at zero lag, by construction a symmetric
matrix. Hence the expression for the learning update is:
\begin{equation}\label{eqn:learn1}
\Delta W \sim \int_{0}^{\infty}\left[C_0 e^{W^{T}t} -e^{W t}C_0
\right]S(t)dt
\end{equation}

The temporal behavior of the weight function has been approximated
by a piece-wise exponential form:
\begin{eqnarray}\label{eqn:STDP}
S(t) = \left\{ \begin{array}{l l} +
e^{t/\tau} & \quad \mbox{$t<0$}\\
0 & \quad \mbox{$t=0$}\\
-e^{-t/\tau} & \quad \mbox{$t>0$}\\
\end{array} \right.
\end{eqnarray}
where $\tau$ is STDP's time-constant, i.e. it expresses the window
over which the plastic changes due to temporal coincidence are
significant. Assuming that the network connections are only
excitatory, and expressing without further loss of generality $W=-I+
A$, we derive the synaptic weight update $\Delta W = \Delta A$ as
follows:
\begin{equation}
\Delta A = -\int_0^{\infty} S(t) e^{W t} C_0 dt + \int_0^{\infty}
S(t) C_0 e^{W^T t} dt
\end{equation}
Given that
\begin{equation}
\int_0^{\infty} e^{-t/\tau} e^{W t} dt  =  \int_0^{\infty}
e^{-t/\tau I + W t} dt =  -\left[W-1/\tau I\right]^{-1}
\end{equation}
We obtain
\begin{equation}
\Delta A = -\left[W-1/\tau I\right]^{-1} C_0 + C_0 \left[W^T-1/\tau
I\right]^{-1}
\end{equation}
Leading finally to
\begin{equation}\label{eqn:learnFinal}
\Delta A \sim \left[I-\frac{\tau}{1+\tau} A\right]^{-1} C_0 - C_0
\left[I-\frac{\tau}{1+\tau} A^{T}\right]^{-1}
\end{equation}
after dropping the multiplying constant $\tau/(1+\tau)$. From this
expression it is possible to derive that the weight update is
anti-symmetric, and that a perfectly symmetric system would not be
modified, as $C_0$ would commute with $A$ (see below, Eq.
\ref{Lyapunov}). Of course, any small initial asymmetry will
eventually be blown up. We can also see that STDP's time constant
also introduces the same multiplying factor $\tau/(1+\tau)$ for $A$,
which can be absorbed by a renormalization; we will assume therefore
$\tau/(1+\tau) \rightarrow 1$ for the remaining of the exposition.
Consistently, the limiting behavior of Eq.~\ref{eqn:learnFinal}
implies $\Delta A(\tau\rightarrow 0)=0$.

\subsection{Minimization of loops and dynamics}\label{SuppInf:Minimization}

Now we can estimate the effect of the synaptic time-dependent
plasticity expressed by Eq.~\ref{eqn:learnFinal} on the topology of
the network. For this, we will postulate a penalty or energy
function for what we will call ``loopiness'' of the network. A
measure of the number of loops occurring in the network can be
obtained by summing the trace of the exponentiation of the network
connectivity matrix, $\sum_k \text{tr}~\left[A^k\right]/k$. This
loop density can be simply minimized by making the connections
vanish, so we need to introduce a regularization penalty to avoid
this effect; an obvious measure of the strength of the connections
in a network is $\text{tr}~\left[A A^{T}\right]$, which in a binary
graph would be equivalent to the total number of links. We postulate
then the following ``loopiness'' energy:
\begin{equation}\label{eqn:loopinessDefine}
\mathcal{E} = \sum_k \frac{1}{k} \text{tr}~\left[A^k\right] -
\frac{1}{2}\text{tr}~\left[A A^{T}\right]
\end{equation}

The change in this energy upon small changes $\Delta A$ is expressed
as $\Delta \mathcal{E} \sim \text{tr}~\left[ \partial_{A}
\mathcal{E} \Delta A^{T} \right]$; it can be easily verified that
$\partial_{A} \mathcal{E} = \left(I-A^{T}\right)^{-1} - A$, and
therefore:
\begin{eqnarray}\label{eqn:loopinesChange1}
\Delta \mathcal{E} &\sim& - \text{tr}~\left[K_1\right] - \text{tr}~\left[K_2\right]\nonumber \\
K_1 &=& (I-A^{T})^{-1} \left[ (I-A)^{-1} C_0 - C_0
(I-A^{T})^{-1} \right]  \nonumber  \\
K_2 &=&  A \left[C_0 (I-A^{T})^{-1}-(I-A)^{-1} C_0\right]
\end{eqnarray}

We will demonstrate in what follows that the traces of $K_1$ and
$K_2$ are strictly semi-positive under the synaptic changes elicited
by STDP (i.e. Eq.~\ref{eqn:learnFinal}), and therefore the loopiness
energy can only decrease over time. We will consider firs that the
correlation matrix can be approximated by the identity, i.e. $C_0
\approx I$. This case can be solved analytically; we will show
numerically that the same conclusions hold for the generic case of
any $C_0$ that results from the dynamical system determined by $A$,
drawn from a Gaussian or Poisson distribution.

Let us consider $\text{tr}~\left[K_1\right]$, rewritten as:
\begin{equation}\label{eqn:term1}
\text{tr}~\left[ (I-A^{T})^{-1}(I-A)^{-1} - (I-A^{T})^{-2} \right]
\end{equation}
and which is of the from
\begin{equation}\label{eqn:term12}
\text{tr}~\left[(PP^{T} - P^2)\right]
\end{equation}
For any matrix $P$,
$$
\text{tr}~\left[(P^{T}-P)(P^{T}-P)^{T}\right] = \text{tr}~\left[R
R^{T}\right]\ge 0
$$
which upon expanding leads to
\begin{equation}\label{eqn:term13}
\text{tr}~\left[PP^{T}-P^2\right]\ge 0
\end{equation}
and ensures the positivity of $\text{tr}~\left[K_1\right]$, under
the assumption $C_0 \approx I$. Moreover, we can show that this is
valid for an arbitrary $C_0$. Under the assumption of stability and
homogeneous Gaussian noise, the correlation and weight matrices are
related by the Lyapunov equation \cite{RISKEN, DELSOLE}:
\begin{equation}\label{Lyapunov}
W C_0 + C_0 W^{T} = -Q Q^{T}
\end{equation}
where $QQ^{T}$ is the generalized temperature tensor of the noise,
whose components $Q_iQ_j=\sigma^2_{ij}$ are the corresponding noise
variances. For the case we are considering, $QQ^T=I$, a symmetric
system $W=W^T$ has the solution $C_0=-W^{-1}/2$. A formal solution
for the general case is \cite{HandJ}
\begin{equation}\label{eqn:lyap_solution}
C_0 = \int_{0}^{\infty} e^{W^T t} e^{W t} dt
\end{equation}
Assuming homogeneous noise, this reduces in our case to
\begin{equation}\label{eqn:term14}
C_0 = \int_{0}^{\infty} e^{-2I t} e^{A^T t} e^{A t} dt
\end{equation}
Following the derivation in \ref{eqn:term12}-\ref{eqn:term13}, the
full expression for $\text{tr}~\left[K_1\right]$ can be written as:
\begin{equation}
\text{tr}~\left[K_1\right] = \text{tr}~\left[ R R^T C_0 \right]
\end{equation}
Given that
\begin{equation}
\text{tr}~\left[ R R^T e^{A^T} e^{A}\right]=\text{tr}~\left[
(e^{A}R)(e^{A}R)^T \right] \ge 0
\end{equation}
It follows through \ref{eqn:term14} that $\text{tr}~\left[K_1\right]
\ge 0$

NEW STARTS HERE @@@@@@@@

Similarly, the second term in Eq.~\ref{eqn:loopinesChange1},
$\text{tr}~\left[K_2\right]$, can be rewritten as
$$
\text{tr}~\left[A^{T}(I-A)^{-1}
C_0\right]-\text{tr}~\left[A(I-A)^{-1} C_0\right]$$ which can be
reduced to
$$
\text{tr}~\left[(A^{T}-A)(I-A)^{-1} C_0\right]
$$
Replacing $A(I-A)^{-1}$ by $(I-A)^{-1}-I$, the term can be
transformed to
\begin{eqnarray}\label{eqn:noneg1}
\text{tr}~\left[C_0\right]-\text{tr}~\left[(I-A^{T})(I-A)^{-1}
C_0\right] = \text{tr}~\left[C_0\right]-\text{tr}~\left[W^{T}W^{-1}
C_0\right]
\end{eqnarray}
Assuming that $QQ^T=I$ in Eq.~\ref{Lyapunov}, and pre-multiplying by
$W^{-1}$, we obtain
\begin{eqnarray}
W^{-1}C_0W^T = -W^{-1}-C_0 \\
\text{tr}~[W^T W^{-1} C_0] = - \text{tr}~[ W^{-1}] - \text{tr}~
[C_0]
\end{eqnarray}
from which we derive:
\begin{equation}\label{eqn:noneg2}
\text{tr}~[K_2] = 2\text{tr}~[C_0] + \text{tr}~[W^{-1}]
\end{equation}
Now we can use the formal solution of the Lyapunov equation,
Eq.~\ref{eqn:lyap_solution}, and the fact that for a stable matrix
such as $W$ (i.e. all its eigenvalues have negative real components)
we can write
$$
W^{-1} = - \int_0^{\infty} e^{W t} dt
$$
and modify Eq.~\ref{eqn:noneg2} accordingly:
\begin{equation}
\text{tr}~[K_2] = 2 \int_0^{\infty}\text{tr}~[ e^{W t}e^{W^T t}] dt
- \int_0^{\infty} \text{tr}~[e^{W t}] dt
\end{equation}
Given that for any matrix, the eigenvalues satisfy
$\lambda(e^W)=e^{\lambda(W)}$, we get
\begin{equation}\label{eqn:noneg3}
\text{tr}~[K_2] =  2 \int_0^{\infty} \sum_{k=1}^{N} e^{2
\text{Re}(\lambda_k) t} dt - \int_0^{\infty} \sum_{k=1}^{N}
e^{\lambda_k t} dt
\end{equation}
where $\lambda_k$ are the $N$ eigenvalues of $W$. Calling
$\lambda_k=-\mu_k+i \gamma_k,~~\mu>0$, the first term in the r.h.s.
above is
$$
 2 \int_0^{\infty} \sum_{k=1}^{N} e^{- 2 \mu_k t} dt =
 \int_0^{\infty} \sum_{k=1}^{N} e^{- \mu_k t} dt
$$
Now we can compare both terms in the r.h.s. of Eq.~\ref{eqn:noneg3}
for each $k$ and $t$: $ e^{- \mu_k t} \ge  e^{- \mu_k t+i \gamma_k
t} $, which leads directly to $\text{tr}~[K_2] \ge 0$ and completes
the proof of the semi-negativity of the changes in the energy
function (Eq.~\ref{eqn:loopinesChange1}).

Interestingly, this result is related to a property of $M$-matrices.
For an $M$-matrix: (1) the off-diagonal elements are semi-negative,
$M_{ij} \le 0~\forall i \ne j$, and (2) is positive stable,
$\text{Re}[\lambda_i(M)]>0~\forall i$. It can be shown that for any
$M$-matrix of dimension $N$ (Th. 5.7.23 in \cite{HandJ}) $
\text{tr}~\left[M^{T}M^{-1}\right] \le \text{tr}~[I] = N $. Choosing
$-W$ as an $M$-matrix, the theorem leads to a similar result for the
non-negativity of tr$[K_2]$ when $C_0$ is close to the identity.

We have assumed throughout that the system is in a regime of
dynamical stability, and presented a case for the stabilizing effect
of STDP by linking loops and eigenvalues in Eq.~\ref{EigLinear}. It
follows that loopiness minimization (Eq.~\ref{eqn:3}) is equivalent
to maximization of stability (as defined by the l.h.s. of
Eq.~\ref{EigLinear}), constrained by the total matrix weight. We can
further understand this by explicitly expanding to first order the
update equation~\ref{eqn:learnFinal} to see the effect on
$\mathcal{U} = - \sum_i \log |\lambda_i|$. Assuming again that
$QQ^T=I$, the solution to the Lyapunov equation
(Eq.~\ref{eqn:lyap_solution}) can be approximated by a series
expansion in powers of $A$ \cite{HandJ}; to first approximation $C_0
\simeq I+\frac{1}{2}(A+A^{T})$, leading to $\Delta A \sim A-A^{T}$.
Through Eq.~\ref{EigLinear} we obtain $\delta \mathcal{U}\simeq
\frac{1}{2} \text{tr}~\left[A ~\delta A + \delta A ~A\right]$, and
in turn $\delta \mathcal{U}\simeq \left(\text{tr}~\left[A^2\right] -
\text{tr}~\left[AA^{T}\right]\right)\leq 0$, making the system more
stable.

UP TO HERE @@@@

\section*{Acknowledgments}
We would like to acknowledge the helpful contributions of Lucas
Monz\'{o}n, University of Colorado at Boulder, and Gustavo
Stolovitzky, IBM Research, to the mathematical analysis.

\bibliographystyle{unsrt}

\bibliography{Kozloski_and_Cecchi_Topological_STDP}

\newpage

\begin{figure}[htb]
\begin{center}
\includegraphics[width=3.5in]{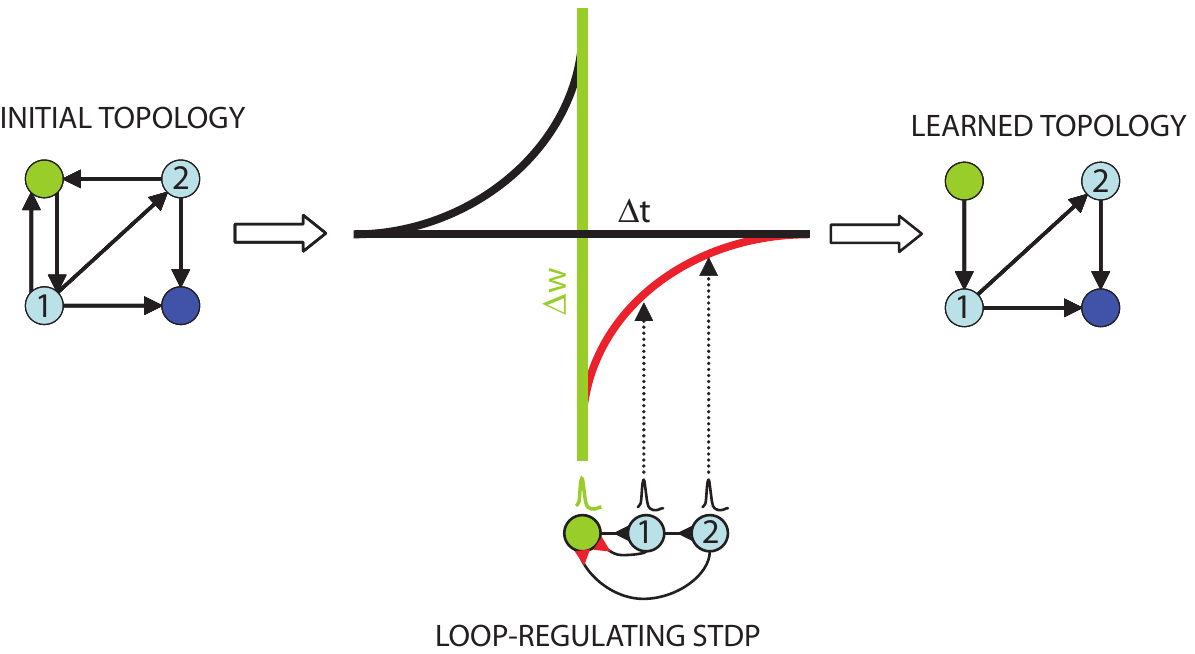}
\caption{Schematic of the topological effect of STDP. Feedback
connections in an initial topology (left) from first ($1$) and
second ($2$) order ``follower'' neurons (light blue) to a
``trigger'' neuron (green) create loops of length $k=2$ and $k=3$.
These connections are selectively penalized by the STDP learning
rule (lower middle, red). The plot (middle) depicts this rule, with
the time difference between follower (black) and trigger (green)
action potentials on the $x$-axis, and the expected synaptic
modification on the $y$-axis. When spikes successfully propagate
through the loopy network they feed back to the trigger, arriving at
the follower-trigger synapse immediately after the trigger neuron
fired and resulting in synaptic depression (red). Through repeated
spike propagation events, STDP results in a completely feed forward
learned topology (right) to the output neuron (dark blue).}
\end{center}
\end{figure}

\newpage
\begin{figure}[htb]
\begin{center}
\includegraphics[width=2.5in]{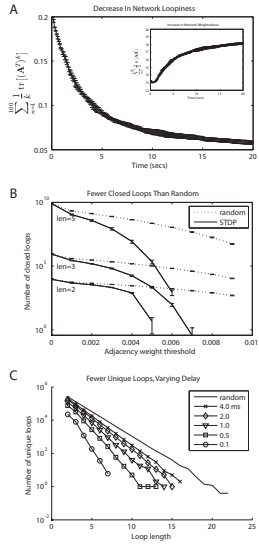}
\caption{Global topological effects of STDP. A) A monotonic decrease
in the loopiness measure (see Equation \ref{eqn:3}, first term) over
time is observed in a simulated network of $100$ neurons undergoing
STDP. Simultaneously, STDP results in a net increase in the
weightedness of the network (inset, see Equation \ref{eqn:3}, second
term). Shown here and in B) is the average of 8 separate simulations
of $20$ seconds of network activity; error bars are standard
deviation. B) Number of closed loops of length $5$, $3$, and $2$,
decreases as a function of weight threshold for network connections.
Dotted lines show counts for randomized networks with same number of
total connections. C) Number of unique loops sampled from five
networks with varying synaptic delays following $10$ seconds of
simulated activity, and from a random network with $5,000$
connections. Number of loops is shown as a function of loop length.
Loops were sampled across different learned networks while
maintaining the number of network connections at $5,000$ by varying
the weight threshold (from $0.003$ to $0.0046$) for each delay.
Greater synaptic delays decreases the loop-eliminating topological
effect of standard STDP.}
\end{center}
\end{figure}

\newpage
\begin{figure}[htb]
\begin{center}
\includegraphics[width=4.5in]{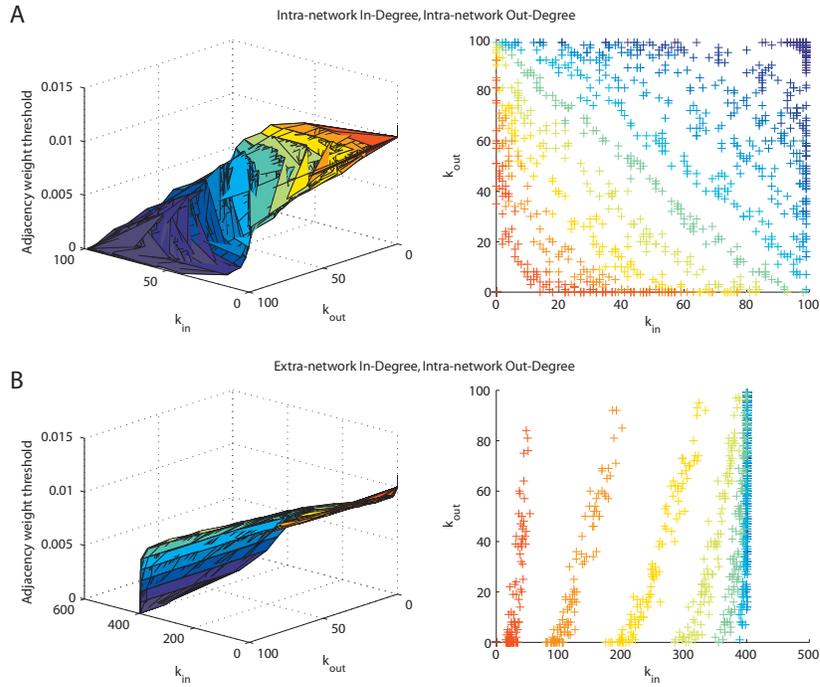}
\caption{Local topological effects of STDP. A) Inverse relationship
of in-degree versus out-degree of intra-network connections for each
neuron in a network after 10 seconds of STDP across multiple weight
thresholds for network connections. Colors in left and right panels
correspond to a weight threshold used to construct the network over
which degrees were measured. The color key can be read from the left
panels' vertical axes and corresponding color found along each
manifold. B) Correlated extra-network in-degree and intra-network
out-degree indicate an opposite effect of STDP on extra-network
inputs.}
\end{center}
\end{figure}

\newpage
\begin{figure}[htb]
\begin{center}
\includegraphics[width=4.5in]{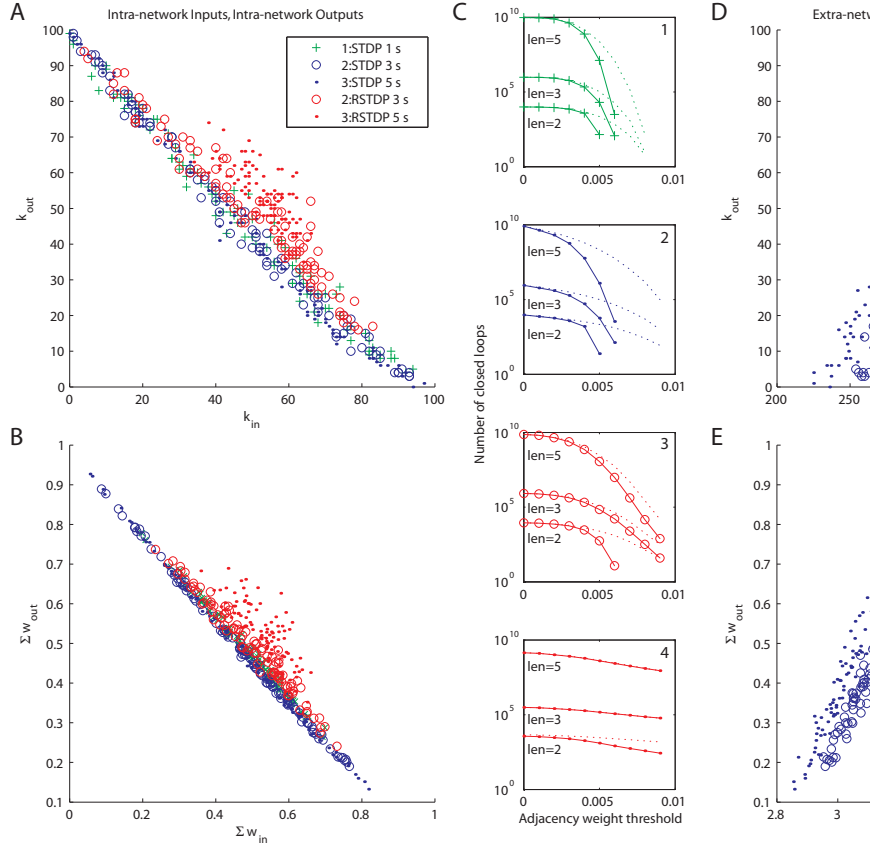}
\caption{Effect of reverse STDP. A) In-degree versus out-degree of
intra-network connections following different durations and
polarities of STDP, shows a strong inverse relationship for standard
STDP. Adjacency weight threshold was $0.005$. Green markers
correspond to the network after $1$ second of standard STDP,
followed by $3$ or $5$ seconds of standard STDP (blue markers) or
reverse STDP (red markers). B) Total synaptic input versus output
for intra-network connections shows a similar inverse relationship
for standard STDP. C) Number of closed loops of length $5$, $3$, and
$2$, is decreased by standard STDP, and restored with reverse STDP
(plotted as in Fig. 2B). D) In-degree of extra-network inputs versus
out-degree of intra-network outputs, plotted as in A, shows a strong
positive correlation for standard STDP. Adjacency weight threshold
was $0.007$ for extra-network inputs and $0.005$ for intra-network
outputs. E) Total synaptic extra-network input versus total synaptic
intra-network output, plotted as in B, shows a similar positive
correlation. In each panel the topological effects of STDP are
reversed by reverse STDP.}
\end{center}
\end{figure}

\newpage
\begin{figure}[htb]
\begin{center}
\includegraphics[width=4.5in]{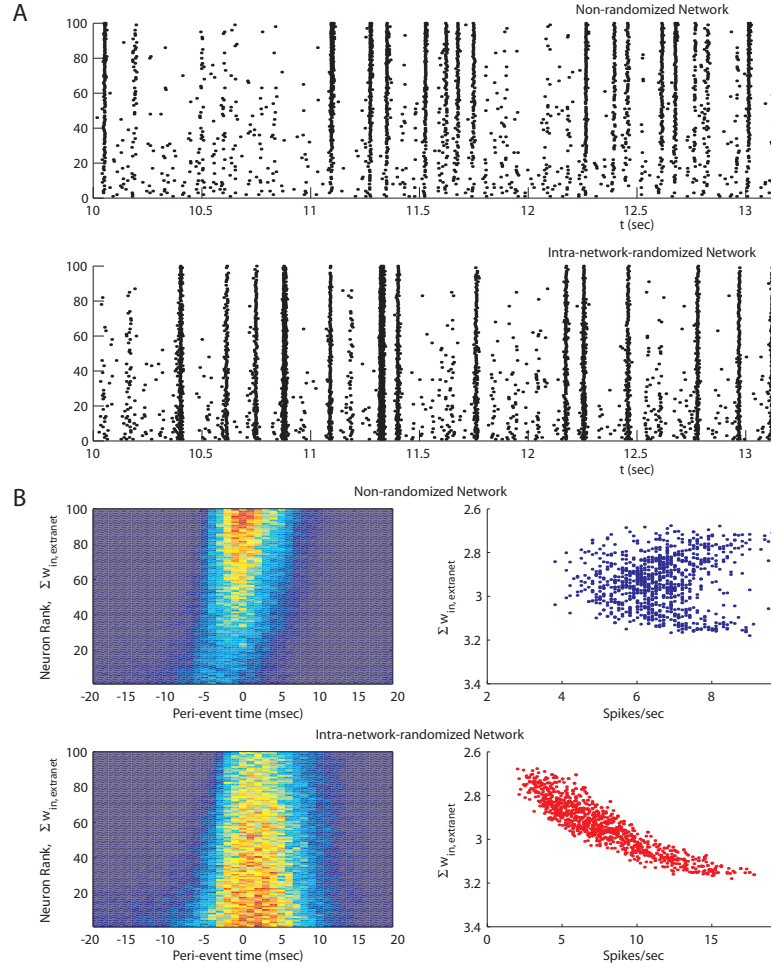}
\caption{Dynamical effects of STDP. A) Raster plot of the spiking
activity for a network after STDP (top), and for a surrogate network
where intra-network weights were reassigned randomly to network
connections, thus destroying STDP-learned topology (bottom). Each
point corresponds to a spike for each neuron. Each neuron was
assigned a rank according to the sum of its extra-network input
weights, with the lowest rank corresponding to the highest sum. B)
Peri-event time histograms for each neuron in the STDP network (top
left) and its surrogate (bottom left), pooled across $8$ separate
simulations (bin width, 2msec). Histograms show different network
propagation properties. Spike counts and extra-network weights for
the same networks do not co-vary in the STDP-learned topology (top
right), but are highly correlated for the surrogate (bottom right).
C) Peri-stimulus time histograms summed across all simulations and
all neurons for the STDP network (blue) and three surrogates, in
which the intra-network connections (red), extra-network connections
(green) or both (magenta) were randomized (top). Skewness versus
kurtosis of these histograms averaged across $8$ separate
simulations each (bottom, error bars show standard deviation)
indicates the network distribution of spikes is more peaked with
more spread for the STDP-learned topology. Inset table shows
P-values of unpaired t-tests of skew (upper right triangle) and
kurtosis (lower left triangle) measurement distributions from each
of $8$ simulations between each of the $4$ conditions, plus the
entire distribution of randomized networks (red, green, magenta,
icons). Squares colored yellow are significant, with stars
indicating P-values' orders of magnitude (from $P<0.05$ to
$P<0.0005$).}
\end{center}
\end{figure}

\renewcommand{\thefigure}{A-\arabic{figure}}
\setcounter{figure}{0}

\end{document}